# Thresholds Optimization for One–Bit Feedback Multi-User Scheduling

Mohammed Hafez**, Ahmed El Shafie††, Mohammed Shaqfeh*, Tamer Khattab†, Hussein Alnuweiri*, Hüseyin Arslan**

†Electrical Engineering, Qatar University (e-mail: tkhattab@ieee.org).
††Electrical Engineering Department, University of Texas at Dallas, USA (e-mail: ahmed.elshafie@utdallas.edu).
*Department of Electrical Engineering, Texas A&M Qatar (TAMUQ) (e-mail: {mohammad.shaqfeh,hussein.alnuweiri}@qatar.tamu.edu).
**Department of Electrical Engineering, University of South Florida, Tampa, USA (e-mail: mhafez@mail.usf.edu and arslan@usf.edu).

*Abstract*—We propose a new one-bit feedback scheme with scheduling decision based on the maximum expected weighted rate. We show the concavity of the 2-user case and provide the optimal solution which achieves the maximum weighted rate of the users. For the general asymmetric $M$-user case, we provide a heuristic method to achieve the maximum expected weighted rate. We show that the sum rate of our proposed scheme is very close to the sum rate of the full channel state information case, which is the upper bound performance.

## I. INTRODUCTION

Temporal and spectral fluctuations are fundamental characteristics of the wireless channel, which occur due to physical phenomena such as fading and shadowing [1]. Consequently, channel-aware adaptive transmission techniques [2] and dynamic resource allocation, using opportunistic scheduling, are applied practically in modern communication systems to maintain good performance under this dynamic environment. The underlying objective of these schemes is exploiting the fading wireless channels when they are at their peak conditions to achieve significant capacity gains [3], [4]. Explicit training sequences (i.e., pilot signals) are used in current wireless communication systems to enable the receivers to measure the instantaneous channel conditions so that coherent detection of the transmitted signals can be applied [5]. In opportunistic scheduling schemes, channel state information (CSI) of all back-logged mobile users in the network should be known at the central scheduler, i.e. base-station. The mobile terminals inform the central scheduler about their CSI using explicit feedback messages. The main drawback is that the CSI feedback consumes a considerable portion of the total air-link resources. Moreover, with the consideration of massive multiple-input multiple-output (MIMO) as an enabling technology for 5G networks, the associated pilot contamination problem rises as a performance limiter [6]–[8]. Consequently, there is a need to develop new transmission technologies with reduced CSI feedback from the users to the base-station. This comes at an advantage to network operators who can release the bandwidth resources that are reserved for CSI feedback to be used for supporting more user data traffic.

The main technical challenge in the design of a reduced-feedback multi-user scheduling scheme is maintaining the prospected capacity gain due to multi-user diversity. This is an important research topic that is seldom investigated thoroughly in the literature [9]. Extensive surveys on feedback reduction methods are provided in [9]. The switched-diversity multi-user scheduling schemes are considered as reduced-feedback schemes [10]–[12]. These schemes are based on one-bit feedback per user per CSI with pre-determined priority order of the users. The CSI feedback is controlled by a threshold on the channel condition to decide the one-bit feedback to be sent to the base-station. The optimization of the users feedback thresholds depends on their assigned priority order.

In this letter, we propose a new approach to the one-bit feedback problem in which the users scheduling is based on the maximum expected reward, i.e., the weighted rate, conditional on the one-bit feedback sent from all users. We optimize the feedback thresholds of each user for this scheme and show that the performance of this limited-feedback scenario is not far from the full-CSI feedback case. We provide analytic solutions for the 2-user case. For the general case of $M$ asymmetric users, we provide a heuristic method to achieve the maximum expected weighted rate.

## II. SYSTEM MODEL

We consider a broadcast channel with a single base-station and $M$ users. We assume a block-fading channel model where the channels remain unchanged during the coherence time and changes from one coherence time duration to another. The channels are independent from one link to another. The source is not aware of the CSI over each individual channel block. However, it knows the statistical information of the channels of all users, which is presented by the probability density function (PDF) of the achievable rate $r_i$, denoted by $f_i(r)$ for the channel of user $i$. A one-bit feedback signal from each user per channel block is used to aid the base-station in selecting which one of the users will be scheduled in a given channel block. The one-bit feedback of a user indicates if its channel in a given channel block is above or below a predetermined threshold of the achievable rate, denoted by $r_i$.

The objective of the scheduling task is to select the user with the highest weighted rate, where the weight of a user, denoted by $\mu_i$ for user $i$, is predetermined before the actual operation (based on quality-of-service (QoS) requirements for each user) and hence is not subject to optimization. Since the source does not have exact knowledge of the CSI of all users, it incorporates its knowledge of the PDFs of the users' channels and the received one-bit feedback from each user in the scheduling decision. Therefore, in every channel block, the base-station schedules the user with the highest weighted *expected achievable rate* conditioned on the received feedback information from that user, $\mu_i \mathbb{E}[R_i(k)|b_1, b_2, \ldots, b_m]$, where $\mathbb{E}[\cdot]$ denotes the expected value of the argument, $R_i(k)$ denotes the achievable rate of user $i$ during the channel block $k$, and $b_i$ is the one-bit feedback from user $i$. For the rest of the letter, the time index $k$ is omitted for simplicity.

### A. Optimization Problem

We optimize the feedback thresholds of the users to maximize the weighted-sum rates of the users, denoted $\Phi$

$$\Phi = \sum_{i=1}^{M} \mu_i \tilde{R}_i \tag{1}$$

where $\tilde{R}_i$ is the expected rate of each user (i.e., averaged across channel blocks). Therefore, we can write the main

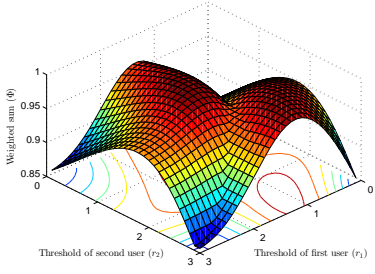

Fig. 1. The change of the weighted rates with the thresholds in the 2-user scenario.

optimization problem as

$$\{r_1^*, \ldots, r_M^*\} = \arg\max_{\{r_1, \ldots, r_M\}} \Phi. \qquad (2)$$

Let us define two quantities for each user, $R_i^+ = \mathbb{E}[R_i | R_i > r_i]$ and $R_i^- = \mathbb{E}[R_i | R_i < r_i]$. These can be obtained using:

$$R_i^+ = \frac{\int_{r_i}^{\infty} r f_i(r)\, dr}{1 - F_i(r_i)}, \quad R_i^- = \frac{\int_0^{r_i} r f_i(r)\, dr}{F_i(r_i)} \qquad (3)$$

where $F_i(r)$ is the cumulative density function (CDF) of the achievable rate of the channel of user $i$.

Moreover, we need to define two quantities to study the effect of the feedback of one user on the scheduling probability of another user, $\Omega_{ij}^+ = \Pr\{\mu_i R_i > \mu_j R_j | R_i > r_i\}$ and $\Omega_{ij}^- = \Pr\{\mu_i R_i > \mu_j R_j | R_i < r_i\}$, where $i \neq j$. These can be obtained using

$$\Omega_{ij}^x = \begin{cases} 1 & \text{if } \mu_j R_j^+ < \mu_i R_i^x \\ F_j(r_j) & \text{if } \mu_j R_j^- < \mu_i R_i^x < \mu_j R_j^+ \\ 0 & \text{if } \mu_i R_i^x < \mu_j R_j^- \end{cases}, \qquad (4)$$

where $x \in \{-, +\}$. Hence, $\tilde{R}_i$ is given by

$$\tilde{R}_i = R_i^+ [1 - F(r_i)] \prod_{j \neq i} \Omega_{ij}^+ + R_i^- F(r_i) \prod_{j \neq i} \Omega_{ij}^-. \qquad (5)$$

## III. OPTIMIZED THRESHOLDS

In this section, we will investigate the problem of finding the optimum thresholds of each user. We start with the 2-user case. Then, we extend the analysis to the general $M$-user case.

### A. 2-User System

Based on the construction of the system, and the structure of (4), we will find that we have 6 different formats of $\Phi$, each of them represents a different combination of the values of (5). Figure 1 illustrates the effect of the selection of the thresholds on the weighted of the expected rates of the users $\Phi$, we can notice that the system has 2 different peaks. Hence, this problem is a non-convex problem.

Consider the case $\mu_1 R_1^+ > \mu_2 R_2^+$. This will limit the number of available formats of $\Phi$ into 3, which are shown in (7). Also, we can rewrite (7c) into (8) and (9).

*Proposition 1:* A local peak for the value of $\Phi$ can be found in the region where the selected thresholds meet one of the following conditions

$$\mu_1 R_1^+ > \mu_2 R_2^+ > \mu_1 R_1^- > \mu_2 R_2^-, \qquad (6a)$$
$$\mu_2 R_2^+ > \mu_1 R_1^+ > \mu_2 R_2^- > \mu_1 R_1^-. \qquad (6b)$$

*Proof:* From (8), it is shown that (7c) is always larger than (7a), as the value of $(\mu_2 R_2^+ - \mu_1 R_1^-)$ is always positive. The is also true for (9) and (7b), where $(\mu_1 R_1^- - \mu_2 R_2^-)$ is always positive. Similarly, we can show that case of $(\mu_1 R_1^+ < \mu_2 R_2^+)$ provides the same results. ∎

Based on Proposition 1, the search for the optimum thresholds has been limited into 2 regions instead of 6.

*Proposition 2:* The optimum thresholds for condition (6a) are given by

$$r_1^* = \frac{\mu_2}{\mu_1} R_2^+, \quad r_2^* = \frac{\mu_1}{\mu_2} R_1^- \qquad (10)$$

and similarly for the case of (6b),

$$r_1^* = \frac{\mu_2}{\mu_1} R_2^-, \quad r_2^* = \frac{\mu_1}{\mu_2} R_1^+. \qquad (11)$$

*Proof:* See Appendix A. ∎

### B. $M$-User System

In the $M$-user case, the exact solution of the optimization problem for finding the global peak is not feasible. Hence, we provide a heuristic method to find the solution. Back to the definition of the function in (4), the number of possible formats as a function of the number of users is $3M!$. Following the same criteria of the 2-user case, we can extend the condition in (6) to

$$\mu_1 R_1^+ > \mu_2 R_2^+ > \cdots > \mu_M R_M^+ > \mu_1 R_1^- > \mu_2 R_2^- > \ldots > \mu_M R_M^-. \qquad (12)$$

Hence, the number of the local peaks would be $M!$. In general, for any number of users $M$, the formula of the rate based on the condition in (12) is given by

$$\tilde{R}_1 = R_1^+ [1 - F_1(r_1)] + R_1^- \prod_{k=1}^{M} F_k(r_k),$$
$$\tilde{R}_i = R_i^+ [1 - F_i(r_i)] \prod_{k=1}^{i-1} F_k(r_k) \qquad \forall i > 1, \qquad (13)$$

hence,

$$\Phi = \sum_{j=1}^{M} \left[ \mu_j R_j^+ [1 - F_j(r_j)] \prod_{n=0}^{j-1} F_n(r_n) \right] + \mu_1 R_1^- \prod_{n=1}^{M} F_n(r_n). \qquad (14)$$

This can be reformulated into $M!$ different formulas as

$$\Phi_m = \sum_{a=0}^{m-1} \left[ \mu_a R_a^+ [1 - F_a(r_a)] \prod_{b=0}^{a-1} F_b(r_b) \right] + \mu_m g_m \prod_{b=0}^{m-1} F_b(r_b)$$
$$+ \sum_{a=m+1}^{M} \left[ (\mu_a R_a^+ - \mu_m R_m^-)[1 - F_a(r_a)] \prod_{b=0}^{a-1} F_b(r_b) \right]$$
$$+ (\mu_1 R_1^- - \mu_m R_m^-) \prod_{b=0}^{M} F_b(r_b) \qquad (15)$$

where $F_0(r_0) = 1$, $\mu_0 = 0$, and $m = 1, 2, \ldots, M!$. Each of these formulas represents a region of thresholds where a local peak exist. To find the optimum thresholds, we equate the first derivative of $\Phi$ with respect to $r$ to zero. That is,

$$\frac{\partial \Phi_m}{\partial r_i^*} = 0, \qquad (16)$$

and based on (13), we will find that optimum thresholds will take the following values:

$$\Phi = \begin{cases} \mu_1 \int_0^\infty r f_1(r) dr & (\mu_2 R_2^+ < \mu_1 R_1^-) & (a) \\ \mu_1 R_1^+ \left[1 - F_1(r_1)\right] + \mu_2 F_1(r_1) \int_0^\infty r f_2(r) dr & (\mu_1 R_1^- < \mu_2 R_2^-) & (b) \\ \mu_1 R_1^+ \left[1 - F_1(r_1)\right] + \mu_1 R_1^- F_1(r_1) F_2(r_2) + \mu_2 R_2^+ F_1(r_1) \left[1 - F_2(r_2)\right] & (\mu_2 R_2^- < \mu_1 R_1^- < \mu_2 R_2^+) & (c) \end{cases} \quad (7)$$

$$\Phi = \mu_1 \int_0^\infty r f_1(r) dr + \left[\mu_2 R_2^+ - \mu_1 R_1^-\right] F_1(r_1) \left[1 - F_2(r_2)\right] \quad (8)$$

$$\Phi = \mu_1 R_1^+ \left[1 - F_1(r_1)\right] + \mu_2 F_1(r_1) \int_0^\infty r f_2(r) dr + \left[\mu_1 R_1^- - \mu_2 R_2^-\right] F_1(r_1) F_2(r_2) \quad (9)$$

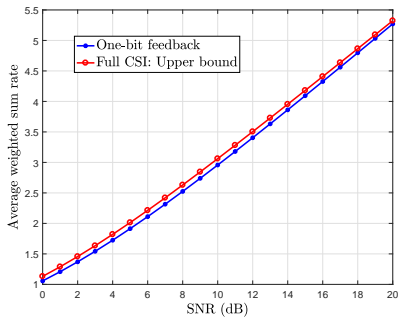

Fig. 2. Comparing the average weighted rate of our proposed one-bit feedback scheme with the full-CSI case (upper bound performance).

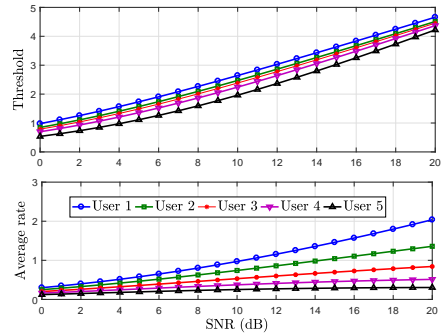

Fig. 3. Change of thresholds and rates with average SNR.

$$r_M^* = \frac{\mu_1}{\mu_M} R_1^-,$$
$$r_{i \neq \{1, M\}}^* = \frac{\mu_{i+1}}{\mu_i} \left[\hat{r}_{i+1}^* F_{i+1}(\hat{r}_{i+1}^*) + R_{i+1}^+ \left(1 + F_{i+1}(\hat{r}_{i+1}^*)\right)\right],$$
$$r_1^* = \frac{\mu_2 \left[\hat{r}_2^* F_2(\hat{r}_2^*) + R_2^+ \left(1 + F_2(\hat{r}_2^*)\right)\right] - \mu_M \hat{r}_M^* \prod_{k=2}^M F_k(\hat{r}_k^*)}{\mu_1 \left[1 + \prod_{k=2}^M F_k(\hat{r}_k^*)\right]}.$$
(17)

It is noticeable that the values of the thresholds recursively depend on each other. Hence, we will have to find their values numerically using numerical methods such as the Bi-section algorithm. In our case, the Bi-section algorithm is used to calculate the optimum thresholds. This algorithm obtains the optimum thresholds for one of the multiple local-optimas. The selection of that local-optima can be realized using one of the following approaches.

*1) Brute-Force Search:* For this case, for each of the $M!$ regions, the optimum thresholds should be calculated, and compare the values of $\Phi$ to get the maximum value. Such approach will require the recursive process for calculating the thresholds to be repeated $M!$ times.

*2) Threshold-Independent Function:* Another approach is to find a certain function $g_i(\mu_i, f_i(r))$ which is independent of the thresholds. Using this way, we will have only $M$ calculations for the value of the function $g_i$, then order the users based on the corresponding value, and finally calculate the optimum thresholds only once.

*3) Random Selection:* An alternative approach is to randomly select one of the $M!$ regions, calculate the optimum thresholds, and decide based on these thresholds. This approach will have minimal calculations requirements, and we will show that it has a negligible performance loss.

IV. NUMERICAL RESULTS

Each channel coefficient is modeled as a complex Gaussian circuitry-symmetric random variable with zero mean and unit variance. We assume that $\mu_1 = 1.1$, $\mu_2 = 1.05$, $\mu_3 = 1$, $\mu_4 = 0.95$, and $\mu_5 = 0.9$. In Fig. 2 shows that the average weighted rate of our proposed one-bit feedback system is comparable with the full-CSI case (upper bound performance).

Fig. 3 shows the change of the thresholds and average rate for each user. We can notice that the rates are close to each other in the low signal-to-noise ratio (SNR) levels, and the gaps between them become wider as the SNR increases. It is noticeable that the QoS requirements affect the way the sum rate is maximized. Users with higher priorities are allocated more frequently which gives them higher rate at better channel condition. On the other hand, users with lower priorities maintain the same rate regardless of the channel condition.

In Fig. 4, we compare the average weighted rate based on the peak selection method. The comparison is between the random selection of any peak and the maximum peak selection. We can find that randomly selecting any local peak, has a negligible effect on the system performance. which can be a good choice to avoid the complexity of finding the global peak.

V. CONCLUSIONS

We showed that the optimum thresholds are recursively dependent. Hence, Bi-section method can be used to obtain their values. The results showed that the weighted rate of our optimized limited feedback system is very close to the one in the case of full-CSI knowledge. We showed that randomly-selecting one of the local maximas, which reduces the complexity of the system, does not have a notable rate loss. The sum-rate loss is less than 2%.

APPENDIX A
OPTIMAL THRESHOLD FOR 2-USER CASE

To find the optimum thresholds for the 2-user case, we solve the following constrained optimization problem:

$$\max_{r_i} . \Phi, \text{ s.t. } 0 \leq r_i, \forall i \in \{1, 2\}. \quad (18)$$

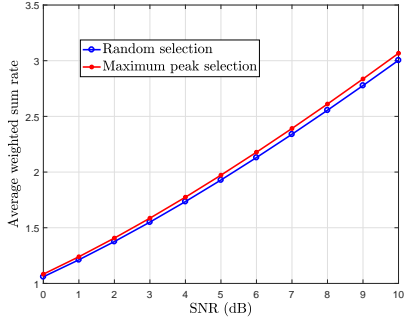

Fig. 4. Comparing two different peak-selection methods.

The maximum of $\Phi$ is obtained using (6) under the constraint that $\mu_2 R_2^- < \mu_1 R_1^- < \mu_2 R_2^+ < \mu_1 R_1^+$. The optimization problem can be then stated as follows:

$$\max_{r_i} . \ \mu_1 \int_0^\infty r f_1(r) dr + \left[\mu_2 R_2^+ - \mu_1 R_1^-\right] F_1(r_1) \left[1 - F_2(r_2)\right],$$
$$\text{s.t. } 0 \leq r_i \forall i \in \{1,2\}, \ \mu_2 R_2^- < \mu_1 R_1^- < \mu_2 R_2^+ < \mu_1 R_1^+. \quad (19)$$

This can be rewritten as

$$\max_{r_i} . \ \left[\mu_2 F_1(r_1) \int_{r_2}^\infty r f_2(r) dr - \mu_1 \left[1 - F_2(r_2)\right] \int_0^{r_1} r f_1(r) dr\right],$$
$$\text{s.t. } 0 \leq r_i \forall i \in \{1,2\}, \ \mu_2 R_2^- < \mu_1 R_1^- < \mu_2 R_2^+ \leq \mu_1 R_1^+. \quad (20)$$

For a fixed (given) $r_2$, the optimization problem is given by

$$\max_{r_i} . \ \left[K_2 F_1(r_1) - K_1 \int_0^{r_1} r f_1(r) dr\right],$$
$$\text{s.t. } 0 \leq r_i \leq 1 \forall i \in \{1,2\}, \ \mu_2 R_2^- < \mu_1 R_1^- < \mu_2 R_2^+ < \mu_1 R_1^+. \quad (21)$$

The first derivative is given by

$$\frac{\delta \Phi}{\delta r_1} = (K_2 - K_1 r_1) f_1(r_1) \quad (22)$$

where $K_2 = \mu_2 \int_{r_2}^\infty r f_2(r) dr$ and $K_1 = \mu_1 \overline{F_2(r_2)}$.[1] If $r_1 > K_2/K_1$, the derivative is negative; hence, the objective function is monotonically decreasing with $r_1$. This means that the optimal solution for a fixed $r_2$ is attained when we set $r_1$ to its lowest feasible value. This value is obtained from the constraint $\mu_2 R_2^- < \mu_1 R_1^- < \mu_2 R_2^+$. If $r_1 < K_2/K_1$, the derivative is positive; hence, the objective function is monotonically increasing with $r_1$. This means that the optimal solution for a fixed $r_2$ is attained when we set $r_1$ to its highest feasible value. This value is obtained from the constraint, $\mu_2 R_2^- < \mu_1 R_1^- < \mu_2 R_2^+$.

The second derivative is given by

$$\frac{\delta^2 \Phi}{\delta r_1^2} = -K_1 f_1(r_1) + (K_2 - K_1 r_1) \frac{\delta f_1(r_1)}{\delta r_1}. \quad (23)$$

If $(K_2 - K_1 r_1) \frac{\delta f_1(r_1)}{\delta r_1} \leq 0$, then the objective function is concave. Setting the first derivative to zero, we get $r_1^* = K_2/K_1$. This condition maintains the concavity of the problem and it is the optimal solution if and only if it satisfies the constraints. We note that $\mu_2 \int_{r_2}^\infty r f_2(r) dr \geq \mu_2 r_2 \overline{F_2(r_2)}$; hence, $r_1^* = K_2/K_1 = \mu_2 \int_{r_2}^\infty r f_2(r) dr / \mu_1 \overline{F_2(r_2)} >$

[1] Throughout this letter, $\overline{\mathcal{X}} = 1 - \mathcal{X}$.

$(\mu_2 r_2 \overline{F_2(r_2)})/(\mu_1 \overline{F_2(r_2)}) = (r_2 \mu_2)/\mu_1$. To sum up, $r_1^* > (\mu_2 r_2)/\mu_1$.

In that case, we can convert the constraint from (6) into three constraints. That is,

$$\mu_2 R_2^+ - \mu_1 R_1^- > 0 \leftrightarrow (\mu_2 R_2^+ - \mu_1 R_1^-) F_1(r_1) \left[1 - F_2(r_2)\right] > 0,$$
$$\mu_1 R_1^- > \mu_2 R_2^-, \ \mu_1 R_1^+ > \mu_2 R_2^+. \quad (24)$$

Next, we check whether or not the optimal solution satisfies the three constraints. Using the constraint, the second term of the objective function is positive and can be bounded as

$$K_2 F_1(r_1) - K_1 \underbrace{\int_0^{r_1} r f_1(r) dr}_{<r_1 F_1(r_1)} > (K_2 - K_1 r_1) F_1(r_1) \geq 0. \quad (25)$$

Thus, we have $r_1 = K_2/K_1$. Hence, the root of the first derivative satisfies the constraint. The second constraint can be rewritten as

$$\gamma_2 = \mu_1 F_2(r_2) \int_0^{r_1} r f_1(r) dr - \mu_2 F_1(r_1) \int_0^{r_2} r f_2(r) dr > 0. \quad (26)$$

We prove this part by contradiction. Assume that we have $\mu_1 r_1^* \geq \mu_2 r_2$ and the constraint is not satisfied. Then,

$$\mu_1 F_2(r_2) \int_0^{r_1} r f_1(r) dr \leq \mu_2 F_1(r_1) \int_0^{r_2} r f_2(r) dr < \mu_2 F_1(r_1) r_2 F_2(r_2). \quad (27)$$

Since $\mu_1 F_2(r_2) \int_0^{r_1} r f_1(r) dr < \mu_1 F_2(r_2) r_1 F_1(r_1)$,

$$\mu_1 F_2(r_2) r_1 F_1(r_1) < \mu_2 F_1(r_1) r_2 F_2(r_2) \leftrightarrow \mu_1 r_1 < \mu_2 r_2. \quad (28)$$

This contradicts the first condition, i.e., $\mu_1 r_1 > \mu_2 r_2$; hence, for the second constraint to be satisfied (to be greater than zero), $r_1$ must equal to $r_1^*$. The third constraint can be shown to be satisfied using the same steps.